# New Method to Calculate the Sign and Relative Strength of Magnetic Interactions in Low-Dimensional Systems on the Basis of Structural Data

**L. M. Volkova[1] and S. A. Polyshchuk[1]**

[1] Institute of Chemistry, Far Eastern Branch of Russian Academy of Science, 690022 Vladivostok, Russia;

***E-mail****: volkova@ich.dvo.ru*

The connection of strength of magnetic interactions and type ordering the magnetic moments with crystal chemical characteristics in low-dimensional magnets is investigated. The new method to calculate the sign and relative strength of magnetic interactions in low-dimensional systems on the basis of the structural data is proposed. This method allows to estimate magnetic interactions not only inside low-dimensional fragments but also between them, and also to predict the possibility of the occurrence of magnetic phase transitions and anomalies of the magnetic interactions. Moreover, it can be used for search of low-dimensional magnets among the compounds whose crystal structures are known. The possibilities of the method are illustrated in an example of research of magnetic interactions in familiar low-dimensional magnets $SrCu_2(BO_3)_2$, $CaCuGe_2O_6$, $CaV_4O_9$, $Cu_2Te_2O_5Cl_2$, $Cu_2Te_2O_5Br_2$, $BaCu_2Si_2O_7$, $BaCu_2Ge_2O_7$, $BaCuSi_2O_6$, $LiCu_2O_2$, and $NaCu_2O_2$.

## 1. INTRODUCTION

Despite intensive studies correlating the magnetic state of the material with its crystal chemical characteristics, the highest achievement in this field still is the discovery of the Goodenough–Kanamori–Anderson rules for the determination of the sign and magnitude of exchange in insulators, which was accomplished as far back as 1950–1960s [1–4]. According to these rules, the linear cation–anion–cation interaction (*M–X–M*) between half-filled orbitals is antiferromagnetic (AF). The cation–anion–cation interaction at less than 90° between half-filled orbitals is ferromagnetic (FM), provided the orbitals are connected with orthogonal anion orbitals. A superexchange including the $\sigma$-bonds is stronger than superexchange for $\pi$-bonds. Among the cations with same electronic configuration, the superexchange of the cations with high valency is stronger.

These rules are widely used; however, the limits of their application are restricted mainly by prediction of ordering type only between the nearest neighbors, which interact through intermediary valent of interactions with intermediate anions. However, possibility to estimate readily the magnetic properties induces many researchers to improve these rules and to approximate to their own objects [5–7].

The dependence of the nearest-neighbor interactions on a bonding angle $M$–$X$–$M$ is obvious. But to predict by these rules a sign and relative strength of magnetic interactions between closely spaced chains is already difficult, and magnetic interactions between clusters, chains or planes positioned at large intervals in low-dimensional magnetic materials is impossible. Whereas there are a sufficient number of compounds in which the magnetic interactions between ions from different structural fragments dominate. Besides, it is difficult to establish by the rules the reason of anomalies of magnetic interactions and magnetic phase transitions in low-dimensional systems, including isomorphic ones. It follows from considering earlier discussion that for crystal chemical factors in definition of a magnetic state of low-dimensional magnetic systems the secondary role was allocated.

The purpose of the present study is the creation of such crystal chemical method, which would be effective for estimation of relative strength of magnetic interactions and a type of ordering the magnetic moments not only inside low-dimensional fragments but also between them and for definition of critical positions of atoms, the insignificant deviations from which can result in the change of a sign or sharp change of magnetic interaction strength.

## 2. DESCRIPTION OF THE NEW CRYSTAL CHEMICAL METHOD TO CALCULATE THE SIGN AND RELATIVE STRENGTH OF MAGNETIC INTERACTIONS

In our work three well-known concepts about the nature of magnetic interactions are used. Firstly, Kramers's idea [8], according to which in exchange couplings between magnetic ions

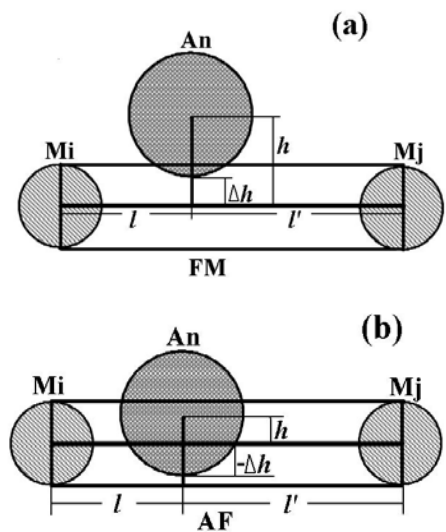

**Fig. 1**. A schematic representation of the intermediate ion $A_n$ arrangement initiated the ferromagnetic (a) and antiferromagnetic (b) interactions between magnetic ions $M_i$ and $M_j$ and parameters ($\Delta h$, $l$, $l'$ and $d(M_i$-$M_j)$), determining sign and strength of magnetic interactions

separated by one or several diamagnetic groups, the electrons of nonmagnetic ions play a considerable role. Secondly, Goodenough–Kanamori–Anderson's model [1–4], in which crystal chemical aspect points clearly to the dependence of strength interaction and the type of orientation of spins of magnetic ions on the arrangement intermediate anions. Thirdly, as in polar Shubin–Vonsovsky's model [9], by consideration of magnetic interactions we took into account not only anions, which are valent bound with the magnetic ions, but also all the intermediate negatively or positively ionized atoms, with the exception of cations of metals with no unpaired electrons.

We have analyzed the connection of magnetic characteristics with the crystal structure in lowdimensional magnets by experimental data given in the literature. The analysis has shown that for definition of ordering type of the magnetic moments and relative strength of interaction between magnetic ions $M_i$ and $M_j$, not having the metal bond, it is necessary to use such parameters as the displacement of intermediate ions $A_n$ from the middle of straight line connecting these magnetic ions; the radii of ion skeletons both magnetic and intermediate ions; the distance between magnetic ions $d(M_i-M_j)$. Besides, it is necessary to assume that the interaction between magnetic ions arises at the moment of crossing the border of space between them by an intermediate ion $A_n$ (Fig. 1). The bounded region of a space between $M_i$ and $M_j$ along a line of their interaction is defined as the cylinder, whose radius is equal to the radius of these magnetic ions. The presence or absence of valent bonds between magnetic and intermediate ions which are included in this space is not taken into account, as differentiated from the Goodenough–Kanamori–Anderson's model.

When some intermediate ions enter into space between two magnetic ions, each of them, depending on its location, aspires to orient the magnetic moments of these ions suitably and make its contribution to occurrence of AF or FM components of magnetic interaction. Let us denote the strength of interaction between magnetic ions $M_i$ and $M_j$, calculated on the basis of the structural data, as $J_{ij}^s$, and contributions to its value, arising under influence of intermediate atoms $A_n$, as $j_n^s$. The sign and value of $J_{ij}^s$ is defined by the sum of these contributions $j_n^s$:

$$J_{ij}^s = \sum_n j_n^s + j^d . \tag{1}$$

If $J_{ij}^s < 0$, the type of ordering of the magnetic moments of $M_i$ and $M_j$ ions is AF; in contrast, when $J_{ij}^s > 0$, the type of the ordering of the magnetic moments is FM.

The value of the contributions $j_n^s$ is defined by expressions:

$$j_n^s = \frac{\Delta h(A_n)\frac{l_n}{l_n'} + \Delta h(A_n)\frac{l_n'}{l_n}}{d(M_i - M_j)^2} \quad \text{(if } l_n'/l_n < 2.0 \text{ ),} \tag{2}$$

and

$$j_n^s = \frac{\Delta h(A_n)\frac{l_n}{l_n'}}{d(M_i - M_j)^2} \quad \text{(if } l_n'/l_n \geq 2.0 \text{ ).} \tag{3}$$

Here $\Delta h(A_n)$ is the difference between the distance $h(A_n)$ from the centre of an ion $A_n$ up to a straight line connecting magnetic ions $M_i$ and $M_j$, and radius ($r_{A_n}$) of an ion $A_n$ (Fig. 1):

$$\Delta h(A_n) = h(A_n) - r_{A_n} . \tag{4}$$

It characterizes a degree of entry of an ion $A_n$ in the space between magnetic ions $M_i$ and $M_j$. If $\Delta h(A_n) < 0$, the ion $A_n$ overlaps (on $|\Delta h|$) the straight line connecting magnetic ions $M_i$ and $M_j$ and initiates the contribution in AF-component of interaction. If $\Delta h(A_n) > 0$, there is a gap (of the width $\Delta h$) in the straight line connecting magnetic ions and ion $A_n$, and this ion initiates the contribution in FM component of the interaction. For calculation $\Delta h(A_n)$, the ionic radii (IR, CN=6) of Shannon [10] ($r_{Cu^{2+}} = 0.73$ Å, $r_{V^{4+}} = 0.58$ Å, $r_{O^{2-}} = 1.40$ Å, $r_{Cl} = 1.81$ Å, $r_{Br} = 1.96$ Å, $r_{B^{3+}} = 0.27$ Å, $r_{Si^{4+}} = 0.40$ Å, $r_{Ge^{4+}} = 0.53$ Å, $r_{Te^{4+}} = 0.97$ Å) are used. Practically identical results turn out when similar size Pauling's ionic radii (CN = 6) are used [11].

$l_n$ and $l_n'$: The lengths of pieces of a straight line $M_i$-$M_j$, connecting magnetic ions are divided by a perpendicular dropped from the centre of the ion $A_n$. The relation $l_n'/l_n$ characterizes a degree of asymmetry of an arrangement of an ion $A_n$ about the middle of the straight line $M_i$-$M_j$. If $l_n'/l_n < 2.0$, the magnetic moments of both ions $M_i$ and $M_j$ undergo the orientation influence of the intermediate ion $A_n$, and the calculation $j_n^s$ needs to be carried out as in Eq. (2). If $l_n'/l_n \geq 2.0$, the ion $A_n$ influences magnetic moment orientation of only nearest magnetic ion and the calculation $j_n^s$ needs to be carried out as in Eq. (3).

$j^d$ - Contribution from direct interaction is taken into account only for magnetic ions located in close proximity, at distances less than two diameters of these ions. We assume that there is some critical distance ($D_c$) between magnetic ions, at which the AF- and the FM contributions from direct interaction are equal and cancel each other, and the deviation from the critical distance towards reduction results in AF coupling and towards increase in FM coupling. The size of this contribution is directly proportional to size of the deviation ($d(M$ - $M)$ - $D_c$) and inversely proportional to the size of skeleton of the magnetic ions ($r_M$) and the distance between them ($d(M$ - $M)$):

$$j^d = \frac{d(M\text{ - }M)\text{ - }D_c}{r_M d(M\text{ - }M)} \qquad (5)$$

Empirically we have established that for interaction between Cu ions $D_c$ must be 2.88 Å.

Each of magnetic ions, as a rule, in the structure is surrounded by several magnetic ions. In each of pairs, which is made up from the ion with neighboring ions, the discrepancy of orientation of the magnetic moments is possible which can result in uncollinear arrangement of the magnetic moments in pair. For final conclusions about the magnetic state of a compound it is necessary in addition to take into account the competition of magnetic interactions and the presence of frustrations in a magnetic system. It is necessary to note that the presence of frustrations of magnetic interactions will not be considered in the present work.

The initial data for calculations of magnetic interactions by this method are only crystallographic parameters, atomic coordinates and ionic radii. Because of this, any errors in definition of the composition or the structures of compounds result in discrepancy of the calculated and experimental parameters. The quantitative results obtained by this method are, certainly, rather rough. However, the sign and the relative strength of magnetic interactions calculated by the method substantially agree with experience.

## 3. CRITICAL POSITIONS OF INTERMEDIATE IONS

It is possible to establish the reasons of occurrence of anomalies of magnetic interactions and magnetic phase transitions in low-dimensional magnets with the help of Eqs. (1) - (3). There are the critical positions of intermediate ions $A_n$, insignificant displacement from which under the influence of temperature, pressure or substitutions can result in a change of the sign or a sharp change of value of the contributions $j_n^s$ in magnetic interaction. In result the sharp change of value $J_{ij}^s$ down to change of ordering type will be observed.

Critical points, where the change $j_n^s$ is possible, are as follows:

(a) $h(A) \approx r_M + r_A$: If the distance $h(A_n)$ from the centre of an ion $A_n$ up to the line of bond $M_i$-$M_j$ between magnetic ions $M_i$ and $M_j$ approaches the sum of radii of ions $M$ and $A_n$ (the ion $A_n$ is near to the surface of the cylinder bounding region of space between of magnetic ions), by insignificant decrease of $h(A_n)$ (displacement of an ion $A_n$ in the region ($h(r_M) \leq r_M + r_{An}$)) arises a strong ferromagnetic interaction between magnetic ions; and by increase $h(A_n)$ (displacement of an ion $A_n$ out of the region ($h(A_n) > r_M + r_{An}$)) this interaction disappears.

(b) $h(A) \approx r_A$ ($\Delta h(A) \approx 0$): If the distance $h(A_n)$ from the centre of an ion $A_n$ up to the line of bond $M_i$-$M_j$ is equal to radius of an ion $A_n$ (ion $A_n$ is near to the line of bond $M_i$-$M_j$), by $h(A_n) = r_{An}$ the interaction between magnetic ions disappear; by insignificant decrease of $h(A_n)$ (overlapping of the line of bond ($h(A_n) < r_{An}$) by an ion $A_n$) arises weak AF interaction, and by increase of $h(A_n)$ (formation a gap between an ion $_{An}$ and the line of bond $M_i$-$M_j$ ($h(A_n) > r_{An}$)) arises a weak FM interaction.

(c) $l_n'/l_n \approx 2.0$ - the insignificant displacement (up to $l_n'/l_n < 2.0$) of an ion $A_n$ to the centre between magnetic ions in parallel by the line connecting $M_i$-$M_j$ results in a sharp increase of strength of interaction.

In the case that there are some intermediate ions $A_n$ between magnetic ions $M_i$ and $M_j$, the following critical points are possible:

(d) When the relation of the sum of the contributions $j_n^s$ in AF-component of interaction to the sum of the contributions $j_n^s$ in FM component of he interaction nears to 1, the interaction between magnetic ions $M_i$ and $M_j$ is weak, and the insignificant displacement of even one of intermediate ions $A_n$ can result in its absolute disappearance or transition of AF-FM.

(e) When even one of the intermediate ions $A_n$ is in critical position such as (a) or (c), the contribution in AF- or FM components of interaction can undergo a sharp changes even from an insignificant displacement of these ions, and can result in a sharp change of strength of the interaction and spin reorientation of magnetic ions.

# 4. THE ILLUSTRATION OF POSSIBILITIES OF THE METHOD ON AN EXAMPLE OF RESEARCH OF MAGNETIC INTERACTIONS IN COMPOUNDS WITH LOW-DIMENSIONAL STRUCTURE

Let us consider the application of the method by the example of calculation of a sign and relative strength of magnetic interactions in well-known low-dimensional magnets, such as: $SrCu_2(BO_3)_2$, $CaCuGe_2O_6$, $CaV_4O_9$, $Cu_2Te_2O_5Cl_2$, $Cu_2Te_2O_5Br_2$, $BaCu_2Si_2O_7$, $BaCu_2Ge_2O_7$, $BaCuSi_2O_6$, $LiCu_2O_2$ and $NaCu_2O_2$. (Table with the results of calculations can be received from the authors of the paper.)

## 4.1. $SrCu_2(BO_3)_2$

The two-dimensional compound $SrCu_2(BO_3)_2$ [12], consists of planes of $CuBO_3$ and Sr atoms between the planes. The nearest-neighbor $Cu^{2+}$ ions form Cu–Cu magnetic dimers (intradimer distances $d$(Cu–Cu)=2.903 Å) arranged in orthogonal dimmer network (interdimer distances $d$(Cu–Cu)=5.133 Å) (Fig. 2a). In the space between Cu ions (cylinder with radius 0.73 Å and in length 2.903 Å), forming the dimer, two ions of oxygen O(1) (Table 1) enter. As these ions locate symmetrically concerning the centre of a line of Cu–Cu bond ($l_n'/l_n = 1$), it is necessary to calculate the contributions in interaction, arising under each ion's action, in Eq. (2). The sum of these contributions is equal to −0.062 Å$^{-1}$. Besides, it is necessary to take into account the contribution from direct interaction of Cu ions $j^d$ [Eq. (5)], as the distance $d$(Cu–Cu) in dimer is less than two diameters of Cu ion (2.92 ° A). This contribution (0.011 Å$^{-1}$) is FM, which reduces as a result of the magnitude of $J_1^s$ up to −0.051 Å$^{-1}$. The type of ordering of the magnetic moments in dimer is AF, as $J_1^s < 0$.

In the space between Cu ions (cylinder with $r$=0.73 Å and $l$=5.133 Å) from various dimers in a plane $ab$, five intermediate ions are placed: ion B, O(1) ion and three O(2) ions. The relation $l_n'/l_n < 2.0$ has only one ion B and ion O(2) (N2); hence, the contributions ($j_B^s$ and $j_{O(2)}^s$) from influence of these ions are calculated in the formula (2), and contributions from other ions in Eq. (3).

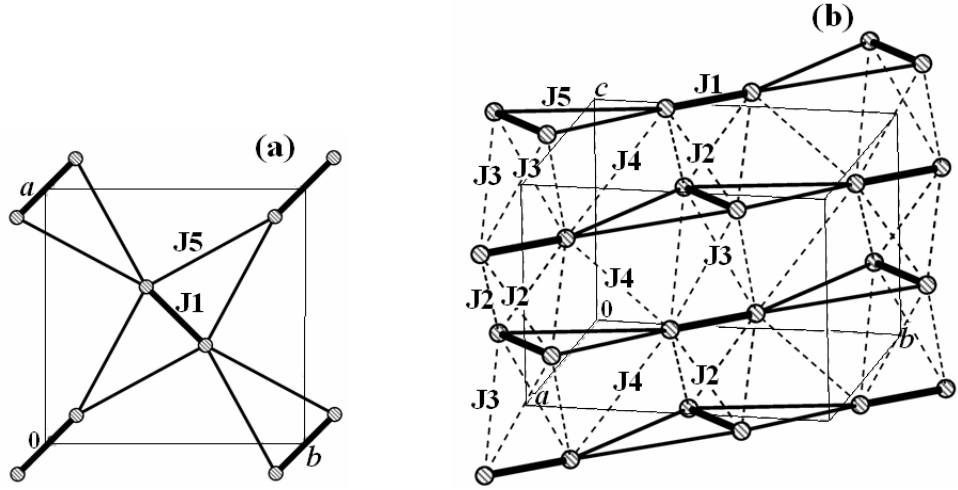

**Fig. 2.** The sublattice of Cu and coupling $J_n$ in SrCu$_2$(BO3)$_2$: (a) $ab$ plane and (b) three-dimensional structure. The thickness of lines shows the magnitude of $J_n$ coupling. AF- and FM couplings are indicated by solid and dashed lines, respectively.

**Table I**. Sign and Strength of Magnetic Interactions ($j_n^s$, $J_n^s$) in SrCu$_2$(BO3)$_2$, Calculated on the Basis of the Structural Data [12]; the Figures Showing an Arrangement of $A_n$ in Space Between Cu Ions, and Parameters $d(Cu$-$Cu)$, $\Delta h$, $l$, и $l'$ Used for Calculations

| $J_n^s$ (Å$^{-1}$) | d(Cu-Cu) and arrangement of $A_n$ | $A_n$ | $h(A_n)$ (Å) | $\Delta h(A_n)$ (Å) | $l$ (Å) | $l'$ (Å) | $j_n^s$ (Å$^{-1}$) | Angles Cu-A$_n$-Cu (°) |
|---|---|---|---|---|---|---|---|---|
| $J_1^s = -0.051$ AF $\sum j_n^s = -0.062$ $j^d = 0.011$ | 2.903 Å | $^a$O(1)X2 | 1.269 | -0.131 | 1.452 | 1.452 | -0.0312 | 97.70 |
| $J_2^s = 0.014$ FM | 3.593 Å | O(1)X2 | 1.546 | 0.146 | 1.152 | 2.441 | 0.0053 | 94.32 |
| | | O(1)X2 | 1.924 | 0.524 | 0.128 | 3.465 | 0.0015 | 64.76 |
| $J_3^s = 0.009$ FM | 4.232 Å | O(1)X2 | 1.717 | 0.317 | 0.876 | 3.356 | 0.0046 | 89.93 |
| | | O(1)X2 | 1.927 | 0.527 | 0.007 | 4.225 | 0.0000 | 65.68 |
| | | O(2)X2 | 1.948 | 0.548 | 0.004 | 4.229 | 0.0000 | 65.38 |
| $J_4^s = 0.009$ FM | 4.796 Å | O(2)X2 | 1.506 | 0.106 | 1.235 | 3.561 | 0.0016 | 106.43 |
| | | O(2)X2 | 1.665 | 0.265 | 1.011 | 3.785 | 0.0031 | 97.53 |
| $J_5^s = -0.037$ AF | 5.133 Å | $^a$1)BX1 | 0.658 | 0.388 | 2.222 | 2.911 | 0.0305 | 150.78 |
| | | $^a$2)O(2)X1 | 0.669 | -0.731 | 1.830 | 3.303 | -0.0655 | 148.47 |
| | | 3)O(1)X1 | 1.049 | -0.351 | 1.618 | 3.515 | -0.0061 | 130.42 |
| | | 4)O(2)X1 | 1.553 | 0.153 | 1.174 | 3.959 | 0.0017 | 105.65 |
| | | 5)O(2)X1 | 1.826 | 0.426 | 0.679 | 4.454 | 0.0025 | 88.09 |

$^a$ The calculation $j_n^s$ is carried out as in Eq. (2), as $l'/l<2.0$. In all other cases the calculation of $j_n^s$ is carried out as in Eq. (3), as $l'/l \geq 2$

The contributions initiated by ion B and by two O(2) ions (in Table I under numbers N4 and N5) have positive values, hence, the contributions are entered in the FM component of interaction, but contributions from ions O(2) (N2) and O(1) (N3) have negative value, hence, the contributions are entered in the AF component. The strength of interdimer magnetic interaction is equal to the sum of these contributions ($J_5^s$ = -0.037 A$^{-1}$). It has negative value as well in a case of intradimer interaction, which point to AF type of spin ordering.

It has been shown before that the values of antiferromagnetic intradimer exchange coupling $J_1$ and interdimer coupling $J_5$ in plane are such that the ratio $J_5/J_1$ is equal 0.60 [13], 0.635 [14] and 0.68 [15]. The relation of these parameters, calculated by us on the basis of the structural data, just as without considering the contribution from direct FM interaction in dimer ($J_5^s/J_1^s$ = 0.60), so taking into account the contribution ($J_5^s/J_1^s$ = 0.72), is in reasonably good agreement with these values.

According to our estimates, inter-layer interactions $J_2^s$ ($d$(Cu–Cu)=3.593 Å), $J_3^s$ ($d$(Cu–Cu)=4.23 Å) and $J_4^s$ ($d$(Cu–Cu)=4.796 Å) (Fig. 2b) are ferromagnetic and are considerably weaker than intra-layer interactions ($J_2^s/J_1^s$ =-0.27, $J_3^s/J_1^s$ = -0.18, $J_4^s/J_1^s$ = -0.18) (Table I), which also confirms conclusions [13–15 ] that $SrCu_2(BO_3)_2$ is a two-dimensional spin system. Notice that the reason to this is not Sr-layers, but absence of the intermediate ions in the central part of space of interaction between Cu ions from various layers.

## 4.2. $CaCuGe_2O_6$

In $CaCuGe_2O_6$ the S=1/2 $Cu^{2+}$ ions are arranged in zigzag chains along the $c$ direction ($d$(Cu–Cu)=3.072 Å) [16]. According to [17–19], despite the fact that from the structural point of view the material has a well-defined

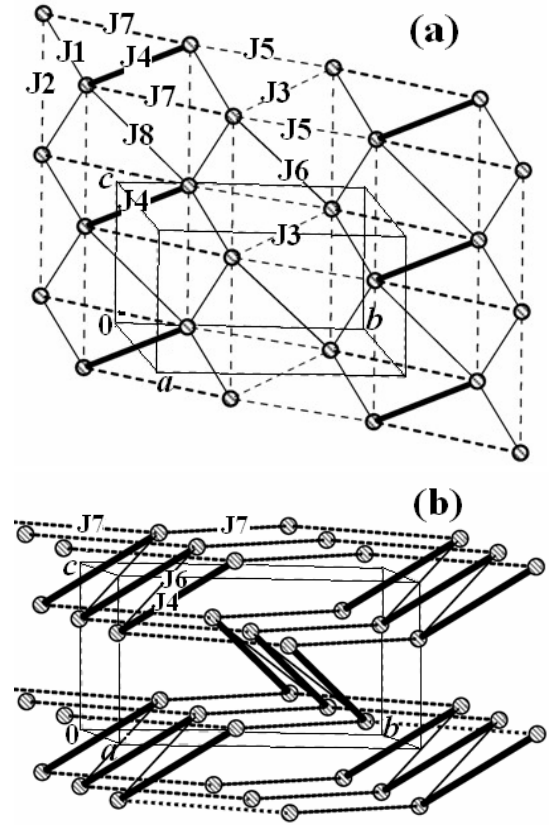

**Fig. 3.** Cu configuration in fragment of $CaCuGe_2O_6$ (a); the strongest interactions 4NN, 6NN and 7NN in $CaCuGe_2O_6$ (b). The thickness of lines shows the size $J_n$ coupling. AF- and FM- couplings are indicated by solid and dashed lines, respectively

one-dimensional arrangement of magnetic ions, the magnetism of $CaCuGe_2O_6$ can be explained by the spin system consisting of antiferromagnetic dimers with weak interdimer interactions. The dimers are assigned to either the third-nearest-neighbor (3NN) or presumably the fourth-nearest-neighbor (4NN) copper pairs [18]. In [19] it is shown that the main role is played by the 3NN[2]. Cu pairs with ferromagnetic (1NN) interdimer couplings.

Our estimation of the spin–spin interactions in $CaCuGe_2O_6$ (Fig. 3a) shows that the anti-ferromagnetic 4NN coupling ($J_4^s$ = -0.051 Å$^{-1}$ (AF), $d$(Cu-Cu) = 6.240 Å) is strongest, but not the 3NN. The nearest-neighor couplings 1NN ($J_1^s$ = -0.003 Å$^{-1}$ (AF), $d$(Cu-Cu) = 3.072 Å), 2NN ($J_2^s$ = 0.001 Å$^{-1}$ (FM), $d$(Cu-Cu) = 5.213 Å) and 3NN ($J_3^s$ = 0.002 Å$^{-1}$ (FM), $d$(Cu-Cu) = 5.576 Å) are really very weak ($J_1^s/J_4^s$ = 0.06, $J_2^s/J_4^s$ = -0.02 and $J_3^s/J_4^s$ = -0.04). In addition, the 1NN and 2NN interaction are unstable. In 1NN interaction there is the possibility of AF→FM transition because of similar values of AF- and FM contributions (critical point ($d$), see Section 3), which arise under the action of O(4) and O(1) intermediate ions, accordingly. Even an insignificant increase $h$ in one of these oxygen ions results in the realization of this transition and increase in the strength of $J_1^s$ coupling. In 2NN interaction, the FM→AF transition is made possible by the displacement of O(1) intermediate ion (the angles Cu-O(1)-Cu = 88.16°), which is located near to a boundary of space between of Cu ions (critical point (a), see Section 3). The increase $h$ of only 0.05 Å of the O(1) ion removes it from the space of interaction and thereby excludes its FM contribution. In result, the $J_2^s$ coupling changes the sign to the opposite and becomes equal to −0.004 Å$^{-1}$.

---

[2]The Cu–Cu distances for 3NN (5.549 ° A) and 4NN (6.213 ° A) are not given precisely in [19]. According to the structural data given in [16] their values are equal to 5.576 and 6.240 ° A, accordingly.

Our consideration was not restricted to four nearest-neighbor couplings and, in addition, estimated four couplings 5NN ($d$(Cu-Cu) = 6.522 Å), 6NN ($d$(Cu-Cu) = 6.550 Å), 7NN ($d$(Cu-Cu) = 7.234 Å) and 8NN ($d$(Cu-Cu) = 7.392 Å). Two of these—5NN ($J_5^s/J_4^s$ = -0.06) and 8NN ($J_8^s/J_4^s$ = 0.09) have also appeared weak, and two others— 7NN ($J_7^s/J_4^s$ = -0.42) and 6NN ($J_6^s/J_4^s$ = 0.16) are much stronger of 1NN, 2NN and 3NN interactions. It seems in the interdimer 7NN and 6NN couplings it is necessary to take into account the description of a magnetic state CaCuGe$_2$O$_6$. Together with the dominant 4NN interactions they form strongly dimerized AF zigzag chains of alternate strong 4NN and weak 6NN interactions along $a$-axis , which are coupled together by rather strong FM interactions 7NN (Fig. 3b).

It is necessary to note that in 7NN coupling, the transition FM→AF is possible by increasing (on 0.06 Å) the value $l$ of an intermediate O(5) ion (critical point (c), see Section 3), and so the size of the relation $l'/l$ becomes less than 2.0. In the result, the contribution in AF component of interaction considerably increases and the value of $J_7^s$ comes nearer to 0.018 Å$^{-1}$.

### 4.3. CaV4O9

The crystal structure of CaV$_4$O$_9$ is determined in two works [20,21]. The unit cell parameters and atomic coordinates of a CaV$_4$O$_9$ differ in these works within the accuracy of the experiment. Hence, the magnetic parameters calculated by the structural data [20,21] are practically equal. In CaV$_4$O$_9$ structure, the magnetic ions V$^{4+}$ are arranged in layers (Fig. 4a), the distance between which is equal to 3.750 Å. The layers consist of two closely spaced planes (1.258 Å) (Fig. 4b). In the planes (Fig. 4c) the ions V$^{4+}$ form the regular squares (so-called larger plaquette) with the sides equal to 3.546 Å (3NN), which are coupled in a square net of distances $d$(V–V)=3.870 Å (4NN). According to our calculations, the AF interactions between V ions in these larger plaquette(the intra-plaquette $J_3^s$= 0.090 Å$^{-1}$) are much stronger than the $J_1^s$ and $J_2^s$ coupling between the planes in the layer ($J_1^s/J_3^s$ = 0.7 and $J_2^s/J_3^s$ = 0.4), despite the shorter interplane distance of V–V (1NN: $d$(V–V)=2.987 Å and 2NN: $d$(V–V)=3.012 Å). These interplane interactions form the goffered net from smaller plaquettes (2NN), which can be presented as four edges of the V$_4$ plane tetrahedron.

The magnetic system CaV$_4$O$_9$ was intensively investigated. For interpretation of the magnetic properties, two alternative models of the systems were used. First, the earlier model with the coupling ratios $J_2 \approx J_1 \approx 2J_3 \approx 2J_4$ considers the interplane couplings ($J_1$ and $J_2$) in a layer as dominant [22,23]. The model is based on the smaller plaquettes (2NN), connected by short linkage (1NN).

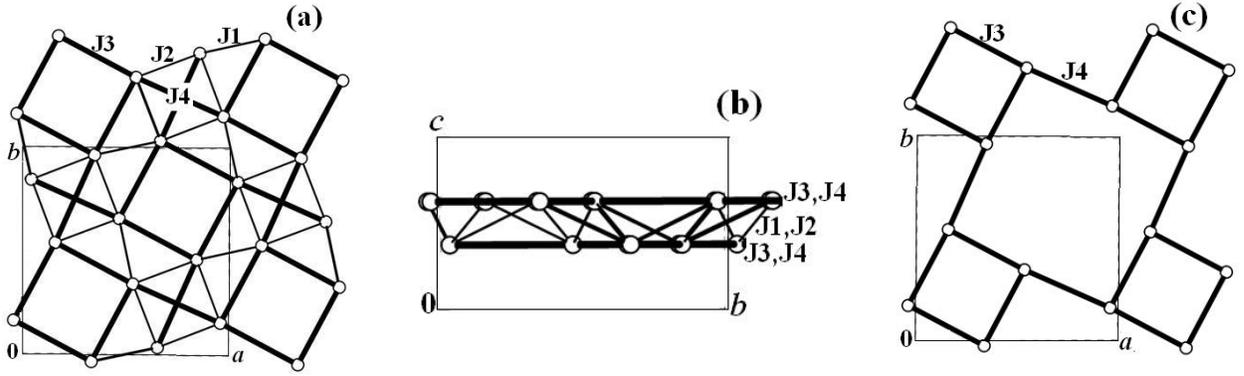

**Fig. 4**. Cu configuration and AF $J_n$ coupling in CaV$_4$O$_9$: the projection of the layer on $ab$-plane (a) and on $bc$-plane, the separate plane of a layer (c).

**Table II**. An Estimate by Various Methods of Relative Strength of Magnetic Couplings in CaV$_4$O$_9$: Inside Small and Large Plaquettes ($J_2/J_3$), Between the Large Plaquettes in a Plane ($J_4/J_3$), Between Planes ($J_1/J_3$, $J_2/J_3$) in a Layer and Between the Intra - Smaller Plaquettes and the Inter- Smaller Plaquettes ($J_1/J_2$).

| Method | $J_2/J_3$ | $J_4/J_3$ | $J_1/J_3$ | $J_1/J_2$ |
|---|---|---|---|---|
| CrCh [given work][a] | 0.44 | 1.10 | 0.70 | 1.59 |
| LDA+U [27][b] | 0.60 | 0.61 | 0.42 | 0.70 |
| LSDA [28][c] | 0.37 | 0.27 | 0.05 | 0.12 |
| SCAD [28][d] | 0.50 | 0.20 | 0.65 | 1.29 |
| Fit [28][e] | 0.65 | 0.26 | 0.68 | 1.03 |
| Neutron [28][f] | 0.49 | 0.12 | 0.49 | 1.00 |
| Neutron [25][f] | 0.39 | 0.09 | 0.39 | 1.00 |

[a]Crystal Chemical method
[b]Local Density Approximation. Modified by a potential correction restoring a proper description of the Coulomb interaction between localized d-electrons of transition metal ions.
[c]Local Spin Density Approximation.
[d]The Self-Consistent Atomic Deformation method.
[e]The Fit results come from fitting the experimental susceptibility.
[f]The couplings deduced from neutron scattering data.

Second, the later model, the interactions in planes of layers are considered as dominant [24–27] (Table II). This model is based on larger plaquettes (3NN) connected by 4NN couplings in square nets in the $ab$ plane. Among the supporters of this model there is no consensus on the strength of the interaction ($J_4$) between the larger plaquettes from one plane. Kodama $et$ $al.$ [24], from the analysis of the neutron inelastic scattering data, have shown that the strongest couplings was within the larger plaquettes ($J3$) and the weakest between the larger plaquettes from one plane ($J_4$), and estimated the interactions as $J_4 \approx 0.1J_3$, $J_1 \approx J_2 \approx 0.4J_3$. From the $ab$-$initio$ calculations, Korotin $et$ $al.$ [26] have found that $J_4$ coupling is smaller than $J_3$, but it is comparable to others exchange couplings $J_4 \approx J_2 \approx 0.6J_3$, $J_1 \approx 0.4J_3$.

The comparison of our data with that of other methods shows (Table II) that the crystal chemical method has estimated $J_2$ and $J_3$ parameters reasonably well. However, the strength of $J_1$ and $J4$ between V ions located on the shortest ($d$(V–V)=2.987 Å) and the longest ($d$(V–V)=3.870 Å) distances is strongly overestimated. As for $J_1$ interaction, the contribution from direct interaction, which is apparently ferromagnetic and decreases the strength of the antiferromagnetic exchange interaction, must be taken into account. The value of $J_4$ couplings can decrease twice, if the contribution from an ion O(3) to divide in 2, because this ion locates in the intersection point of the spaces of two $J_4$ interactions. However, we do not have enough experimental facts to confirm the assumptions.

Thus, our calculations of sign and relative strength of magnetic interactions are consistent with the second model [24–27], which describs the magnetic system CaV$_4$O$_9$ on the basis of larger plaquettes ($J_2^s = 0.4 \, J_3^s$). Besides, they confirm the result of [26], which is indicative of the existence a rather strong $J_4^s$ coupling ($J_4^s = 1.1 \, J_3^s$) between the large plaque plaquettes located in one plane, and weakened coupling between the plaquettes from different planes of one layer ($J_2^s = 0.6 \, J_1^s = 0.4 \, J_3^s$).

## 4.4. Cu$_2$Te$_2$O$_5$Cl$_2$ and Cu$_2$Te$_2$O$_5$Br$_2$

Up to now, there is no general agreement among researchers as to the magnetic state of the isostructural Cu$_2$Te$_2$O$_5$Cl$_2$ [28] and Cu$_2$Te$_2$O$_5$Br$_2$ [28] compounds, which contain clusters of $S = \frac{1}{2}$ Cu$^{2+}$, representing the tetrahedra compressed along $c$-axis with four short edges ($d$(Cu–Cu)=3.230 Å) and two longer edges ($d$(Cu–Cu)=3.591 Å) (Fig. 5a). These tetrahedra provided the basis for the models of a magnetic state of the systems. In the beginning, the tetrahedra were considered weakly coupled units of four spins with couplings $J_1$ and $J_2$ and the ratio of intradimer couplings $J_2/J_1 = 1$ for the chloride (bromide) system [28], which has been refined into $J_2/J_1 \approx 0.66$ for the bromide and $J_2/J_1 < 0.66$ for the chloride systems [29]. In later researches, it was shown that there are strong inter-tetrahedral couplings along the $c$-axis [29,30] and that both the intertetrahedral couplings perpendicular and parallel to the $c$-axis play an important role in ground state order and leads to phase transitions and geometrical frustration [30–36] researches, it was shown that there are strong inter-tetrahedral couplings along the $c$-axis [29,30] and that both the intertetrahedral couplings perpendicular and parallel to the $c$-axis play an important role in ground state order and leads to phase transitions and geometrical frustration [30–36]. Furthermore, in [31] it is suggested that the systems are three-dimensional rather than one-dimensional.

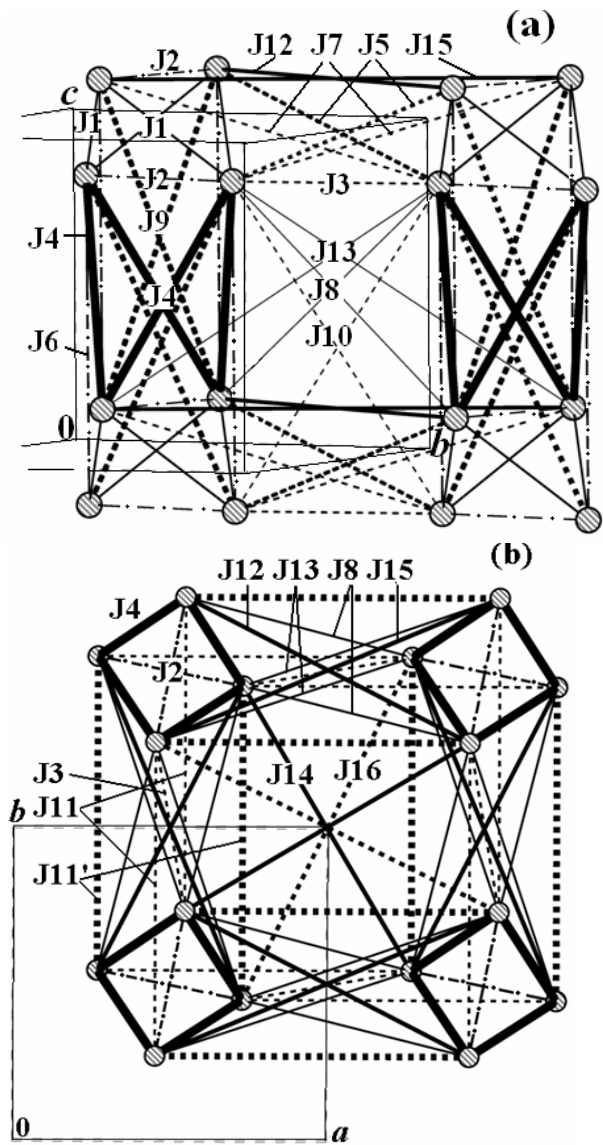

**Fig. 5.** (a) The sublattice of Cu and $J_n$ coupling in Cu$_2$Te$_2$O$_5$Cl$_2$. (b) Projection on the *ab* plane of a layer from coupled tetramers (stretched tetrahedra). The thickness of lines shows the strength of $J_n$ coupling. AF- and FM couplings are indicated by solid and dashed lines, respectively. The possible FM→AF transitions are shown a stroke by dashed lines. The coupling $J_8^s$ and $J_{13}^s$ are FM and $J_6^s$ is AF in Cu$_2$Te$_2$O$_5$Br$_2$, as against Cu$_2$Te$_2$O$_5$Cl$_2$, where $J_8^s$ and $J_{13}^s$ are AF and $J_6^s$ is FM.

Our calculations show that the absolute size of $J_2/J_1$ ratio is closer to 1 (0.95) in Br than in the Cl systems (0.54), and confirm the results of [29], but as distinct [29], the $J_2$ is FM ordering. However, even insignificant decrease (on ~0.03 Å) in $h$ of the intermediate atom O(1) (angle CuO(1)Cu=103.08º (102.41º)) results in a change of the sign of the $J_2$ on opposite and the F→AF transition in both systems (critical point (d), see Section 3). The inter-tetrahedral AF and FM magnetic interactions are much stronger than the intra-tetrahedral interactions (Fig. 5a and b). The strongest AF coupling was the inter-tetrahedral coupling $J_4^s$ ($d$(Cu-Cu)=5.015(5.059) Å, $J_4^s$ = -0.046(-0.047 Å$^{-1}$) in Cl(Br) -systems) in tubes along the $c$-direction. It is possible to present the tubes as alternating along the $c$-axis the edge-sharing tetrahedral of two types: compressed, which we considered yet, and stretched ($d$(Cu-Cu) = 5.015(5.059) Å X4 and $d$(Cu-Cu) = 3.591(3.543) Å X2 in Cl(Br) systems), whose long edges are taken as couplings between the compressed tetrahedra. It is likely that on the basis of the model, it should be considered not the compressed but stretched tetrahedra, as $J_1^s/J_4^s$ =0.22(0.19), and it to be represented as tetramers because the intra-tetrahedral couplings $|J_4^s| >> |J_2^s|$ ($J_2^s/J_4^s$ = -0.13 (-0.17)).

In both Cl(Br) systems these tetramers are coupled in layers, which are perpendicular to $c$ axes, by strong AF interactions $J_{12}^s$, $J_{15}^s$ and $J_{14}^s$ ($J_{12}^s/J_4^s$ = 0.62(0.49), $J_{15}^s/J_4^s$ = 0.58(0.46), and $J_{14}^s/J_4^s$ = 0.37(0.36)), and also by strong $J_{11}^s$ and $J_{11'}^s$ ($J_{11}^s/J_4^s$ = -0.85(-1.06), $J_{11'}^s/J_4^s$ =-0.91(-0.83)) and weak $J_3^s$ and $J_{16}^s$ ($J_3^s/J_4^s$ = -0.26(-0.15) and $J_{16}^s/J_4^s$ = -0.37(-0.08)) FM interactions. Notice that the absolute value $J_{11}^s$ decreases ($J_{11}^s/J_4^s$ = -0.26 (-0.55)) if the large contribution to the interactions from the intermediate ions Cl(1) for Cl compound and O(1) for Br compound, located in critical position near to border of the space of interaction

(critical point (a), see Section 3) are not to take into account.

In addition, between the tetramers in a layer exist the couplings $J_8^s$ and $J_{13}^s$, which are different for Cl and Br systems. In Cl system, the couplings are weak AF ($J_8^s/J_4^s$ = 0.06, $J_{13}^s/J_4^s$ = 0.06), and in Br system, the opposite, i.e. strong FM ($J_8^s/J_4^s$ = -0.47, $J_{13}^s/J_4^s$ = -0.30). This is because the intermediate ions O(3) are located in critical position (critical point (a), see Section 3).

The layers from tetramers are bound together by strong and weak FM interactions $J_9^s$, $J_7^s$, $J_5^s$ and $J_{10}^s$ ($J_9^s/J_4^s$ = -0.87(-0.81), $J_7^s/J_4^s$ = -0.87(-0.89), $J_5^s/J_4^s$ = -0.52(-0.66) and $J_{10}^s/J_4^s$ = -0.11(-0.13)) and weak AF interactions $J_1^s$ ($J_1^s/J_4^s$ = 0.22(0.19), which are similar in both the systems. Notice that the value $|J_7^s/J_4^s|$ decreases till −0.26 (−0.25) if the calculation of $J_7^s$ coupling does not take into account the contribution from an intermediate ion Cu, located in critical position (critical point (a), see Section 3). However, the inter-layer couplings $J_6^s$ (along $c$-axis, $d$(Cu–Cu)=$c$) differ sharply for each of systems. The coupling $J_6^s$ is strong FM ($J_6^s/J_4^s$ = −0.61) in Cl system, and in Br system, opposite, weak AF ($J_6^s/J_4^s$ = 0.23). The reason of difference, is similar existing in the couplings $J_8^s$ and $J_{13}^s$: critical position of O(3) ions. In $J_6^s$ couplings of Cl system there can be a transition F→AF with the decreasing in two times the strength of the coupling at increase (on 0.05 Å) $h$ of O(3) ion. In $J_6^s$ of Br system, opposite there can be a transition AF→F with the increasing of the strength of the coupling at reduction $h$ of O(3) ion on 0.06 Å.

Apparently, the phase transitions to an AF ordered state at TN=18.2 and 11.4 K for Cl and Br systems, respectively [30], are caused by insignificant displacement of O(3) ions at decreasing of temperature. The magnetic ordering transitions FM→AF can take place in $J_6^s$ coupling ($J_6^s$ = 0.028 Å$^{-1}$ (FM) → $J_6^s$ = -0.011 Å$^{-1}$ (AF)) in Cl system, and in $J_8^s$ ($J_8^s$ = 0.022 Å$^{-1}$ (FM) → $J_8^s$ = -0.002 Å$^{-1}$ (AF)) and $J_{13}^s$ ($J_{13}^s$ = 0.014 Å$^{-1}$ (FM) → $J_{13}^s$ = -0.007 Å$^{-1}$ (AF)) couplings in Br system. The opportunity of realization the magnetic ordering transition FM→AF in $J_2^s$ is shown earlier.

Thus, the calculation of the coupling parameters in Cu$_2$Te$_2$O$_5$Cl$_2$ and Cu$_2$Te$_2$O$_5$Br$_2$ on the basis of the structural data shows that the systems are three dimensional. They represent strongly connected tetramers. A structural basis of these tetramers is stretched, instead of compressed tetrahedra, as was considered earlier.

## 4.5. BaCu$_2$Si$_2$O$_7$ and BaCu$_2$Ge$_2$O$_7$

The isomorphic compounds of BaCu$_2$Si$_2$O$_7$ [37–39] and BaCu$_2$Ge$_2$O$_7$ [37,38] contain the chains of Cu2+ ions, running along the $c$-axis. It is experimentally proved [37,38,40–42] that both compounds should be considered as guasi-one-dimensional systems with dominant strong intrachain antiferromagnetic exchange interactions. The intrachain interactions $J_1$ are much larger for the Ge compound ($J_1^{Ge}/J_1^{Si}$ = 2.1; $J_1^{Si}$ = 24.1 meV, $J_1^{Ge}$ = 50 meV) [37,38,40 42]. According to our calculations, the intrachain interactions $J_1$ are also AF (Fig. 6a and b), however, the difference in the magnitude of $J_1$ coupling in Si and Ge compounds is smaller ($J_1^{Ge}/J_1^{Si}$ = 1.3; $J_1^s$ = -0.058 (-0.078) Å$^{-1}$ (AF) in Si(Ge) compound). The interchain interactions are much weaker the intrachain interactions $J_1$. The following values for BaCu$_2$Si$_2$O$_7$ were obtained in [42]: $Jx$ = −0.46 meV (FM) ($|J_x/J_1|$ = 0.02), $J_y$ = 0.20 meV (AF) ($|J_y/J_1|$ = 0.01) and $J_3$ = 0.15 meV (AF) ($|J_3/J_1|$ = 0.006), where $J_x$, $J_y$ and $J_3$ interchain exchange constants along the $a$- and $b$-axes and [1 1 0] directions. This interchain couplings by us are designated as follows (Fig. 6c): $J_x$ as $J_2^s$; $J_y$ as $J_7^s$ and $J_8^s$; $J_3$ as $J_{11}^s$ and $J_{13}^s$.

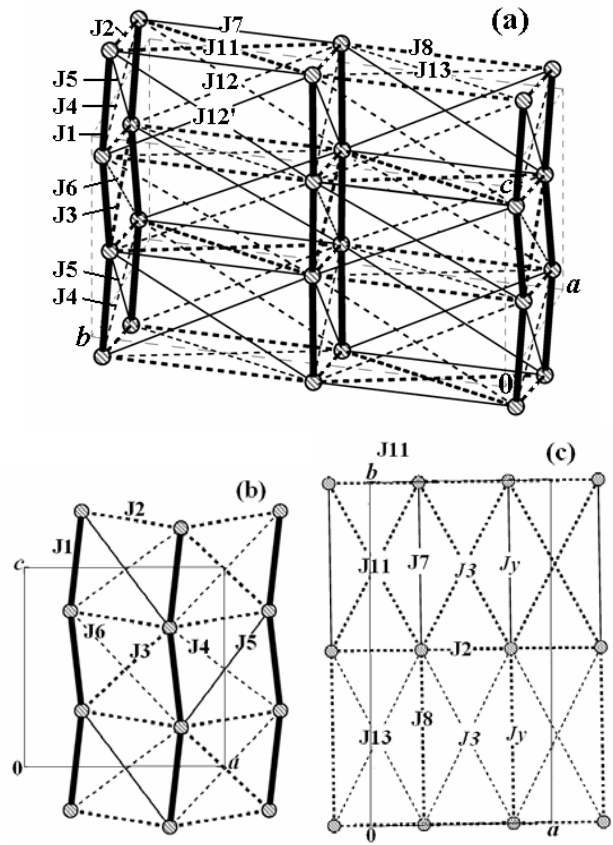

**Fig. 6.** (a) The sublattice of Cu and $J_n$ coupling in powder BaCu$_2$Si$_2$O$_7$. The projections on the $ac$-plane (b) and $ab$-plane (c). The coupling $J_4^s$ in single crystal BaCu$_2$Si$_2$O$_7$ and powder BaCu$_2$Ge$_2$O$_7$ is AF, as differentiated from powder BaCu$_2$Si$_2$O$_7$, which is FM. AF- and FM couplings are indicated by solid and dashed lines, respectively.

Our estimate interchain interactions supports the conclusion [42] that along an $x$-axis in Si compound there is ferromagnetic spin ordering, however, overestimates ($|J_2^s/J_1^s|$ = 0.28) the value of $J_2^s$ ($J_x$) coupling (d(Cu-Cu) = 3.480 Å, $J_2^s$ = 0.016 Å$^{-1}$ (FM)), as the values of others interchain interactions. We found that along a $y$-axis there exist not one $J_y$, but two nonequivalent interactions, one of two $J_7^s$ (d(Cu-Cu) = 6.451 Å, $J_7^s$ = -0.004 Å$^{-1}$ (AF), $|J_7^s/J_1^s|$ = 0.07) is weak antiferromagnetic, and other $J_8^s$ (d(Cu-Cu) = 6.725 Å, $J_8^s = 0.020$ Å$^{-1}$ (FM), $|J_8^s/J_1^s| = 0.34$), opposite, is stronger ferromagnetic. Along direction [1 1 0] also there are not one, but two nonequivalent interactions: $J_{11}^s$ (d(Cu-Cu) = 7.330 Å, $J_{11}^s = 0.016$ Å$^{-1}$ (FM), $|J_{11}^s/J_1^s| = 0.28$) and $J_{13}^s$ (d(Cu-Cu) = 7.572 Å, $J_{13}^s = 0.002$ Å$^{-1}$ (FM), $|J_{13}^s/J_1^s| = -0.03$), which have appeared FM, instead of AF, in contrast to results in [42].

In [38] it is shown that a small interchain coupling also changes with the composition from ferromagnetic at the Si side to antiferromagnetic at the Ge side. Our results confirm this conclusion, however, the distinction between Si and Ge compounds present not in the $J_2$ couplings, which in the both compounds are ferromagnetic, according to our calculation, but in the diagonal couplings $J_4$ (d(Cu–Cu)=4.776 (4.895) Å in Si(Ge) compounds) in $ac$-plane. In Si compound this interaction is weak ferromagnetic ($J_4^s = 0.001$ Å$^{-1}$ (FM), $J_4^s/J_1^s =$ -0.02), and in Ge compound, it is opposite, weak antiferromagnetic ($J_4^s = -0.006$ Å$^{-1}$ (AF), $J_4^s/J_1^s = 0.08$).

The earlier parameters of magnetic interactions $J_n^s$ in $BaCu_2Si_2O_7$ and $BaCu_2Ge_2O_7$, we calculated by the structural data received with powder X-ray diffraction experiments in [37,38]. Apparently, in the powder samples Si and Ge compounds, which were used in these works for definition of the crystal structure and magnetic properties, there are vacancies (~5 %) in a positions of ions oxygen, as ~10 % of Cu ions are a univalent. This assumption stems from the fact that the bond-valence sum of Cu ions (BVS) [43] in them is less than 2 (BVS=1.90 (1.89) for Si(Ge) compounds). The additional argument in favour of this assumption, there are the anomalies of thermal parameters ($B_{iso}$) of oxygen ions in powder samples of Si and Ge compounds [37,38]: negative value of ($B_{iso}$) for atom O(1), and too low value of ($B_{iso}$) for atom O(4). It should be particularly emphasized that in the single crystal sample of $BaCu_2Si_2O_7$ [39], probably, there are no vacancies in the position of oxygen atoms, as BVS of Cu is equal 2.00.

We have calculated and compared the sign and the strength of magnetic interactions $J_n^s$ in powder [37,38] and single crystal [39] samples. Although the maximal difference in values of the $h(A_n)$ distances is 0.1 Å in Si and Ge compounds, all the $J_n^s$ values, except for two ($J_1^s$ and $J_4^s$), do not change the sign and differ on size in limits ~|0.002| Å$^{-1}$. The strength of magnetic interaction $J_1^s$ in the chain along $c$-axis in the single crystal sample has sharply increased (up to −0.070 Å$^{-1}$ (AF)) relative to value in the powder sample ($J_1^s = -0.058$ Å$^{-1}$ (AF)) through the increasing (on |0.011| Å$^{-1}$) of the contribution in the AF component of interaction because of decrease (on 0.07 Å) $h$ of the intermediate O(4) ion. The magnetic interaction $J_4^s$ in single crystal sample remained the same weak ($J_4^s = -0.0004$ Å$^{-1}$ (AF)), as in the powder sample ($J_4^s = 0.0005$ Å$^{-1}$(FM)), but its sign has changed. Notice that the interaction $J_4^s$ is in a critical state (critical point (d), see Section 3), as the relation of the contributions sum in AF- and a FM component of this interaction nearers to 1. The insignificant displacement even of one of intermediate ions $A_n$ can result in the transition AF–FM or to a state, when one type of the contribution can completely cancel another. The $J_4^s$ interaction in single crystal sample is antiferromagnetic at the expense of the increase (on |0.0012| Å$^{-1}$) of the contribution in the AF component of interaction also because of reduction $h$ of intermediate ions O(4), but in this case not one, and two. By this, the contribution in FM component of $J_4^s$ interaction, arising under action of two ions O(2), remains practically constant (0.0062–0.0064 Å$^{-1}$) not only in powder and single crystal samples of Si compounds, but in Ge compound also. From here follows what to receive all spectrum of states $J_4^s$ interactions: weak AF, weak FM and the cancellation of AF interactions by FM interactions are possible not only at the substitution of Si on Ge in these compounds, but also at the creation of vacancies in a positions of oxygen ions. In addition, in $BaCu_2Ge_2O_7$ the $J_4^s$ interaction has one more the critical point (c) (see Section 3), as the relation $l''/l$ of an ion O(4) is equal to 2.0. The strength of this interaction can colossally increase (up to −0.053 Å⁻¹ (AF)), if the will be a displacement (only 0.028 Å) the O(4) ions to the centre between Cu atoms in parallel by straight line, which connects them. In result, the Ge compound transforms from one- to two-dimensional magnetic state at the expense of strengthening AF interchain interactions $J_4$ in $ac$-plane. Whether there is a probability to achieve, it remains by an enigma.

### 4.6. $BaCuSi_2O_6$

The $Cu^{2+}$ ions in $BaCuSi_2O_6$ are arranged in bilayers parallel to the (0 0 1) crystallographic plane (Fig. 7). The intra-dimer, inter-dimer nearest-neighbor within the bilayers and inter-bilayer exchange couplings have been estimated as 4.45, 0.58, and 0.116 meV, respectively [44,45].

Our calculations carried out on the basis of structural data of $BaCuSi_2O_6$ [46] as well as the researches on a basis of high field magnetization data [44,45] show, that AF intradimer interaction $J_1$ ($J_1^s$ = -0.068 Å¹, (Cu-Cu)=2.728 Å) is dominant. The interaction $J_1^s$ is formed by two contributions: antiferromagnetic (−0.076 Å⁻¹), arising from the direct interaction of the Cu ions, and ferromagnetic (0.008 Å⁻¹) induced by the influence of eight intermediate O(1) ions. The interdimer interactions within the bilayer on distances equal lattice constants $a$ and $b$ ($d$(Cu–Cu)=7.042 Å), have appeared nonequivalent. According to our estimates, one of two interactions ($J_3^s$) is the weak AF interaction ($J_3^s / J_1^s$ = 0.08), as well as in [44,45], where $J_3/J_1$ = 0.13, and another ($J_{3'}^s$), opposite, is the strong FM interaction ($J_3^s / J_1^s$ = −0.74). These interactions alternate within the bilayers (Fig. 7). Leaving by O(2) ion the space of $J_{3'}^s$-interaction (displacement of only ~0.1 Å) will reduce the strength of the $J_{3'}^s$-interaction in 20 times (critical point (a), see Section 3). The diagonal interaction $J_5^s$ ($d$(Cu-Cu)= 7.522 Å)) between planes within the bilayer is FM ($J_5^s / J_1^s$ = -0.35). Both of AF inter-bilayer couplings $J_2^s$ ($d$(Cu-Cu)=5.732 Å) and $J_4^s$ ($d$(Cu-Cu)=7.469 Å) are approximately equal and weak interactions ($J_2^s / J_1^s$ = 0.12, $J_4^s / J_1^s$ = 0.15). They are ~5 times weaker than FM interdimer nearest-neighbor within the bilayers $J_{3'}^s$-interactions ($J_2^s / J_{3'}^s$ = -0.16, $J_4^s / J_{3'}^s$ = -0.20) and more two times weaker than diagonal within the bilayers FM $J_5^s$ interactions ($J_2^s / J_5^s$ =-0.34, $J_4^s / J_5^s$ =-0.42). The anti-ferromagnetic interaction $J_4^s$ can increase the strength a three times, if the intermediate O(2) ion will leave the space of interaction, having displaced all on 0.06 Å (critical point (a), see Section 3).

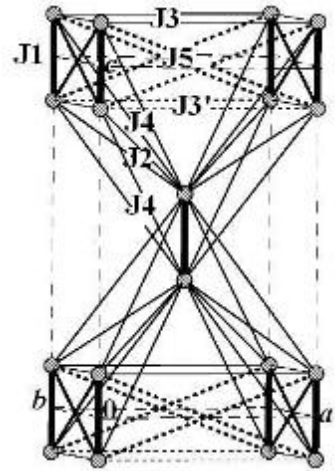

**Fig. 7.** The sublattice of Cu and $J_n$ coupling in $BaCuSi_2O_6$. AF- and FM-couplings are indicated by solid and dashed lines, respectively.

Thus, the estimate of a magnetic state of BaCuSi$_2$O$_6$ by the crystal chemical method qualitatively coincides with conclusions in [44,45]. In addition to [44,45], the existence of nonequivalent AF and FM interdimer within the bilayers couplings is established. The possibility of decrease of the strength of FM interdimer within the bilayers $J^s_{3'}$-couplings and increase AF inter-bilayer $J^s_4$ couplings under influence of temperature or pressure because of displacement of O(2) ions, located in a critical positions near to borders of space of interaction, is shown.

## 4.7. LiCu$_2$O$_2$ and NaCu$_2$O$_2$

Now, by description of a magnetic state of isostructural compounds LiCu$_2$O$_2$ and NaCu$_2$O$_2$ the model, proposed by Drechsler *et al.* [47] on the basis of structural arguments and LDA calculations of LiCu$_2$O$_2$ occupies the leading position. According to this model [47], in LiCu$_2$O$_2$ the dominant interaction is the $J_4$ AF interaction in a linear chain along the *b*-axis (*d*(Cu–Cu)=2*b*=5.717 Å)) (Fig. 8). The nearest-neighbor $J_2$ interactions in this chain are almost twice weaker $J_4$ ($J_2/J_4$= −0.56) and are ferromagnetic. From the point of view of the authors [47], it is the cause of FM-AF frustration in single-chain. Besides, it is found that the interchain interactions $J_1$ (*d*(Cu–Cu)=3.083 Å) and $J_\perp$ (*d*(Cu–Cu)=5.726 Å) are AF and weak ($J_1/J_4$ =0.03 and $J_\perp/J_4$ =0.39). The $J_1$ interactions within a chain pair are especially weak.

This model have confirmed by neutron diffraction measurements on NaCu$_2$O$_2$ [48], according to which $J_4 \gg |J_2| \gg J_1 \sim 0$. In this compound in addition the magnetic couplings in a linear chain up to a distance *d*=4*b* were considered and it was established, that whereas $J_{d \geq 2b}$ couplings are always antiferromagnetic, $J_2$ can be either ferromagnetic or antiferromagnetic. The couplings $|J_{2(d=b)}|$, $J_{3b}$ and $J_{4b}$ are significantly smaller than $|J_4|$.

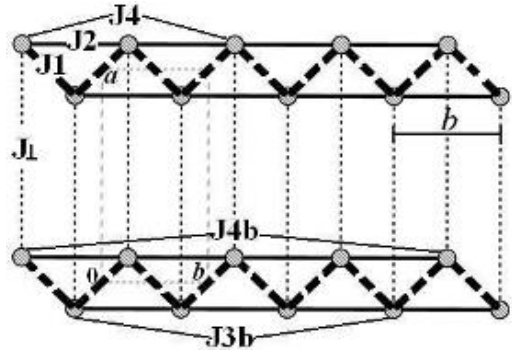

**Fig. 8.** The scheme of interactions between magnetic Cu$^{2+}$ ions in LiCu$_2$O$_2$.

We found, that magnetic interactions in LiCu$_2$O$_2$ [49] and NaCu$_2$O$_2$ [50] considerably differ, in spite of the fact that these compounds are isostructural. According to our estimate, the $J_4$ interaction is strong AF interaction in both compounds. In NaCu$_2$O$_2$ it really dominates over all interactions ($J_4$= −0.045 Å$^{-1}$ (AF), $J_1/J_4$= −0.41, $J_2/J_4$ =0.66, $J_1/J_4$= −0.05, $J_{3b}/J_4$ =0.30 and $J_{4b}/J_4$ =0.41), that confirms the results [48]. However, in LiCu$_2$O$_2$ the $J_4$ interaction ($J_4$= −0.046 Å$^{-1}$ (AF)) is twice as weak as the interchain interaction $J_1$ ($J_1$ = 0.090 Å$^{-1}$ (FM)), though it is much stronger than all other interactions ($J_2/J_4$ =0.58, $J_\perp/J_4$ = −0.03, $J_{3b}/J_4$ =0.25 and $J_{4b}/J_4$ =0.38). The main discrepancies between our results and model [47] consist in the following:

(i) The nearest-neighbor interaction in a linear chain $J_2$ is AF, and interchain interaction $J_1$ within a chain pair is FM on the contrary of [47].

(ii) The interchain interaction $J_1$ is dominant in LiCu$_2$O$_2$ ($|J_2/J_1|$ = 0.30), that confirms results of Masuda *et al.* ([51], model 2). In NaCu$_2$O$_2$ the interaction $J_1$ only in 2.4 times is weaker, than dominant interaction $J_4$, instead of in 36 times, as in model [47].

Despite of these discrepancies (i and ii), our results do not contradict to the conclusions [47] about existence of the geometrical frustration in *single* chains. However, from our point of view, a source of the frustration in *single* chains is the competition $J_{2(d=b)}$, $J_{4(d=2b)}$, $J_{3b}$ and $J_{4b}$ *antiferromagnetic* interactions, instead of the $J_2$ *ferromagnetic* and the $J_4$ *antiferromagnetic* interactions, as it is considered in [47]. Besides, the competition of the inchain *antiferromagnetic* interactions ($J_{2(d=b)}$, $J_{4(d=2b)}$, $J_{3b}$ and $J_{4b}$) and the interchain *ferromagnetic* $J_1$ interactions within a chain pair can be the cause of one more frustration. This frustration in double chains of LiCu$_2$O$_2$ was found earlier [52] and was explained by a competition $J_1$ and $J_2$ *antiferromagnetic* interactions. However, the frustration scenario [52] is in doubt [48,53].

It is significant, that the bond-valence sum of Cu ions, calculated according to [43], are less than 2 (BVS=1.90 in LiCu$_2$O$_2$ [49] and BVS=1.78 in NaCu$_2$O$_2$ [50]) in samples, which structural data were used for calculation of the sign and strength of magnetic interactions. Apparently, the vacancies (up to ~5%) in oxygen positions and in consequence a loss of the magnetic moment at a part (up to ~10%) of Cu ions are characteristic for these compounds. It should result in some divergence of results of researches on various samples.

Thus, the results of calculations show that the offered method correctly defines the dimension of the magnetic subsystem. However, the estimate of strength of magnetic interactions is rough. The discrepancy between the results obtained by the use of our method and other methods is observed mainly when the intermediate ions are in critical positions near to a surface of the cylinder bounded region of a space between magnetic ions (critical point (a), see Section 3), and is expressed in over-estimate of the contributions in a FM component of interaction. Insignificant changes of positions of these atoms, which can be connected to distinction of conditions (temperature, pressure), at which the structure and magnetic properties was investigated, can introduce large mistake into calculations. Therefore, it is necessary to consider two values of strength of magnetic interaction: with the account and without the account of the contributions from intermediate ions located in these positions.

The magnetic properties are extremely sensitive not only to small changes in positions of intermediate ions, but also to presence of vacancies in structure. The advantage of crystal chemical method over others consists in opportunity simply and quickly to estimate the magnetic interactions between ions removed from each other on any distances, knowing only crystal structure of low-dimensional compounds and sizes of ions.

## 5. CONCLUSION

The analysis of connection of magnetic characteristics with the crystal structure in low-dimensional magnets, by experimental data presented in the literature, has shown that the structural factor plays a crucial role in the determining the magnetic state of low-dimensional compounds.

The strength and sign of interaction between magnetic ions are determined by the sum of the contributions in AF- and FM components of magnetic interactions arising under the effect of intermediate ions and depend on such parameters: the displacement of intermediate ions concerning the middle of line, connecting these magnetic ions, the radii both magnetic and intermediate ions and the distance between magnetic ions. The magnitude of the contributions in AF- and FM components of interactions should be maximal, if the intermediate ions are located in narrow space on middle of distances between magnetic ions, but are not near to them. For the maximal contribution to an AF component of interaction the intermediate ions should be approached to an axis, and in a FM component of interaction, on the contrary, to a surface of the cylinder bounding region of space between of magnetic ions. If the magnetic ions are located on close distances, which are less than two diameters of these ions, the contribution from direct interaction between them is taken into account. The critical values of crystal chemical parameters, the insignificant deviation from which is accompanied by sharp change of strength of magnetic interactions or spin reorientations, are established.

The mathematical expression is deduced and the program for calculation of a sign and relative strength of magnetic interactions on the basis of the structural data is developed.

The magnetic interactions in low-dimensional systems $SrCu_2(BO_3)_2$, $CaCuGe_2O_6$, $CaV_4O_9$, $Cu_2Te_2O_5Cl_2$, $Cu_2Te_2O_5Br_2$, $BaCu_2Si_2O_7$, $BaCu_2Ge_2O_7$, $BaCuSi_2O_6$, $LiCu_2O_2$ and $NaCu_2O_2$ are investigated with this method. The results of calculations show that the offered method basically correctly estimates the magnetic interactions and allows to define not only dimension of a magnetic subsystem proceeding from structural data of compounds but also to predict an opportunity of occurrence of magnetic anomalies or change of type of magnetic ordering. It can be used in creation of models for interpretation of experimental results and also for search new low-dimensional magnets.

# ACKNOWLEDGMENTS

We would like to thank Dr. D.Yu. Popov for a development of the program "MagInter" for calculation of a sign and relative strength of magnetic interactions on the basis of the structural data. This work was supported by grant of Far East. Br. Russ. Ac. Sci.